\documentclass{article}

\usepackage{PRIMEarxiv}

\usepackage[utf8]{inputenc} 
\usepackage[T1]{fontenc}    
\usepackage{hyperref}       
\usepackage{url}            
\usepackage{booktabs}       
\usepackage{amsfonts}       
\usepackage{nicefrac}       
\usepackage{microtype}      
\usepackage{lipsum}
\usepackage{fancyhdr}       
\usepackage{graphicx}       
\graphicspath{{media/}}     
\usepackage{amsfonts}  
\usepackage{graphicx}
\usepackage{caption}
\usepackage{subcaption}
\usepackage{float}
\usepackage{amsmath}
\usepackage{adjustbox}
\usepackage{threeparttable}
\usepackage{float}
\usepackage{comment}
\title{HelixDesign-Antibody: A Scalable Production-Grade Platform for Antibody Design Built on HelixFold3
}

\author{
  Jie Gao, Jing Hu, Shanzhuo Zhang, Kunrui Zhu, Sheng Qian, Yueyang Huang, Xiaonan Zhang, Xiaomin Fang\thanks{Corresponding author. Email: fangxiaomin01@baidu.com } \\
  PaddleHelix Team, Baidu Inc. \\
}

\begin{document}
\maketitle

\begin{abstract}

Antibody engineering is essential for developing therapeutics and advancing biomedical research. Traditional discovery methods often rely on time-consuming and resource-intensive experimental screening. To enhance and streamline this process, we introduce a production-grade, high-throughput platform built on HelixFold3, HelixDesign-Antibody, which utilizes the high-accuracy structure prediction model, HelixFold3. The platform facilitates the large-scale generation of antibody candidate sequences and evaluates their interaction with antigens. Integrated high-performance computing (HPC) support enables high-throughput screening, addressing challenges such as fragmented toolchains and high computational demands. Validation on multiple antigens showcases the platform’s ability to generate diverse and high-quality antibodies, confirming a scaling law where exploring larger sequence spaces increases the likelihood of identifying optimal binders. This platform provides a seamless, accessible solution for large-scale antibody design and is available via the \href{https://paddlehelix.baidu.com/app/all/helixdesign-antibody/forecast}{antibody design page of PaddleHelix platform}.
\end{abstract}


\section{Introduction}
Antibody engineering has become a fundamental strategy in the development of therapeutic molecules, diagnostic tools, and research reagents. Through structure-based rational design, antibodies can be optimized to enhance their interaction with antigens, including improvements in binding affinity, thereby enhancing their functional efficacy across a broad range of applications.

Recent advances in high-accuracy structural prediction models, such as AlphaFold-Multimer \cite{evans2021protein}, AlphaFold3 \cite{abramson2024accurate}, HelixFold-Multimer \cite{gao2024precise} and \href{https://paddlehelix.baidu.com/app/all/helixfold3/forecast}{HelixFold3} \cite{liu2024technical}, have shown strong capabilities in differentiating binding from non-binding partners in protein–protein interactions. This has enabled the integration of structure-based design with predictive modeling to drive antibody optimization. In this workflow, given the backbone conformation of an antigen and a reference antibody, inverse folding models like ESM-IF \cite{hsu2022learning} and ProteinMPNN \cite{dauparas2022robust} are used to generate candidate antibody sequences expected to adopt the desired structural configuration. These candidates are then assessed using structural prediction models, which verify whether the antibody binds to the intended epitope and provide interface-level confidence scores. This evaluation helps filter out unstable or misfolded designs by identifying plausible binding conformations. Predicted complex structures are further analyzed for steric clashes or unintended shifts in binding pose. Variants exhibiting realistic and stable binding modes proceed to binding energy analysis using tools such as FoldX \cite{schymkowitz2005foldx} and PRODIGY \cite{xue2016prodigy}, which enable quantitative ranking based on thermodynamic stability and interaction strength.

Despite the effectiveness of this structure-based approach, several challenges persist. Building a complete design pipeline often requires integrating diverse tools with incompatible input formats, parameter conventions, and software environments—an undertaking that can be overwhelming for scientists without programming experience. Selecting appropriate parameters for each tool further compounds the complexity, especially when clear guidelines are lacking. Additionally, structural prediction models remain computationally intensive, typically requiring powerful GPUs to achieve usable throughput. In low-throughput settings, insufficient sampling and filtering can lead to suboptimal design results. Empirical observations in protein design suggest that broader candidate evaluation substantially increases the likelihood of discovering high-quality antibodies.

To address these limitations, we introduce a scalable production-grade binder design platform built upon HelixFold3, HelixDesign-Antibody, which streamlines sequence generation, structural validation, and binding evaluation into a cohesive and scalable workflow. We have validated the effectiveness of this platform on several antigens, demonstrating its ability to generate diverse antibody candidates and reliably predict high-quality complex structures. We also confirm the presence of a scaling law in structure-based design, where broader sequence exploration substantially improves the chances of identifying optimal antibodies.

Key features of the HelixDesign-Antibody are summarized as follows:
\begin{itemize}
\item \textbf{Precise Structural Evaluation via HelixFold3:}
Powered by the HelixFold3 model, the platform provides atomic-level evaluation of antibody–antigen interactions, enabling the quick identification of favorable binding conformations early in the design process. HelixFold3 has been specifically optimized for antigen–antibody scenarios, enhancing the accuracy of structural prediction and scoring.

\item \textbf{High-Throughput Design and Evaluation:}
HelixDesign-Antibody leverages Baidu AI Cloud’s high-performance computing (HPC) platform to enable large-scale generation and evaluation of antibody candidates. This computational capability facilitates efficient exploration of the antibody sequence space and increases the likelihood of identifying designs with high structural confidence and strong predicted binding affinity.

\item \textbf{Multi-Dimensional Scoring and Filtering:}
The platform employs a comprehensive scoring framework encompassing:
(i) sequence-based metrics that reflect evolutionary fitness,
(ii) structure-based assessments of interface geometry and confidence,
(iii) physicochemical evaluations of properties such as hydrophobicity and electrostatics.
This integrative approach facilitates balanced candidate prioritization across sequence, structural, and biophysical dimensions.

\item \textbf{Integrated and User-Friendly Platform:}
Designed with accessibility in mind, the system minimizes coding requirements and configuration complexity, making structure-based antibody engineering more approachable for researchers without software development expertise. Users can explore and experience the full workflow through the \href{https://paddlehelix.baidu.com/app/all/helixdesign-antibody/forecast}{PaddleHelix platform}, which provides an accessible online interface for structure-based antibody design and optimization.
\end{itemize}

\section{Method}

\begin{figure}
\centering
\includegraphics[width=1.0\linewidth]{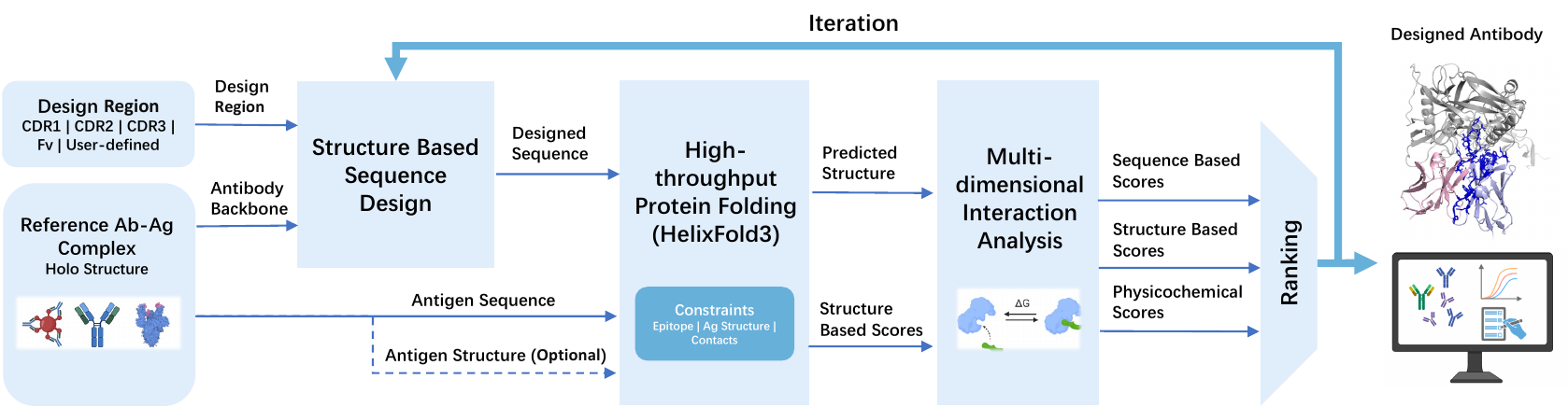}
\caption{HelixDesign-Antibody is a structure-based antibody design pipeline that takes as input the complex structure of a reference antibody–antigen pair. Users can specify the antibody design regions, such as CDR1–3 or the full Fv region, based on their objectives. Guided by structural context, the workflow first generates a set of candidate antibody sequences using inverse folding algorithm. These candidates are then evaluated through high-throughput structure prediction of antibody–antigen complexes using the high-precision folding model HelixFold3. Finally, the predicted structures are assessed using a multi-dimensional scoring system to systematically prioritize antibody candidates to select the high potential ones.}
\label{fig:framework}
\end{figure}

\begin{figure}[htbp]
    \centering
    \includegraphics[width=\linewidth]{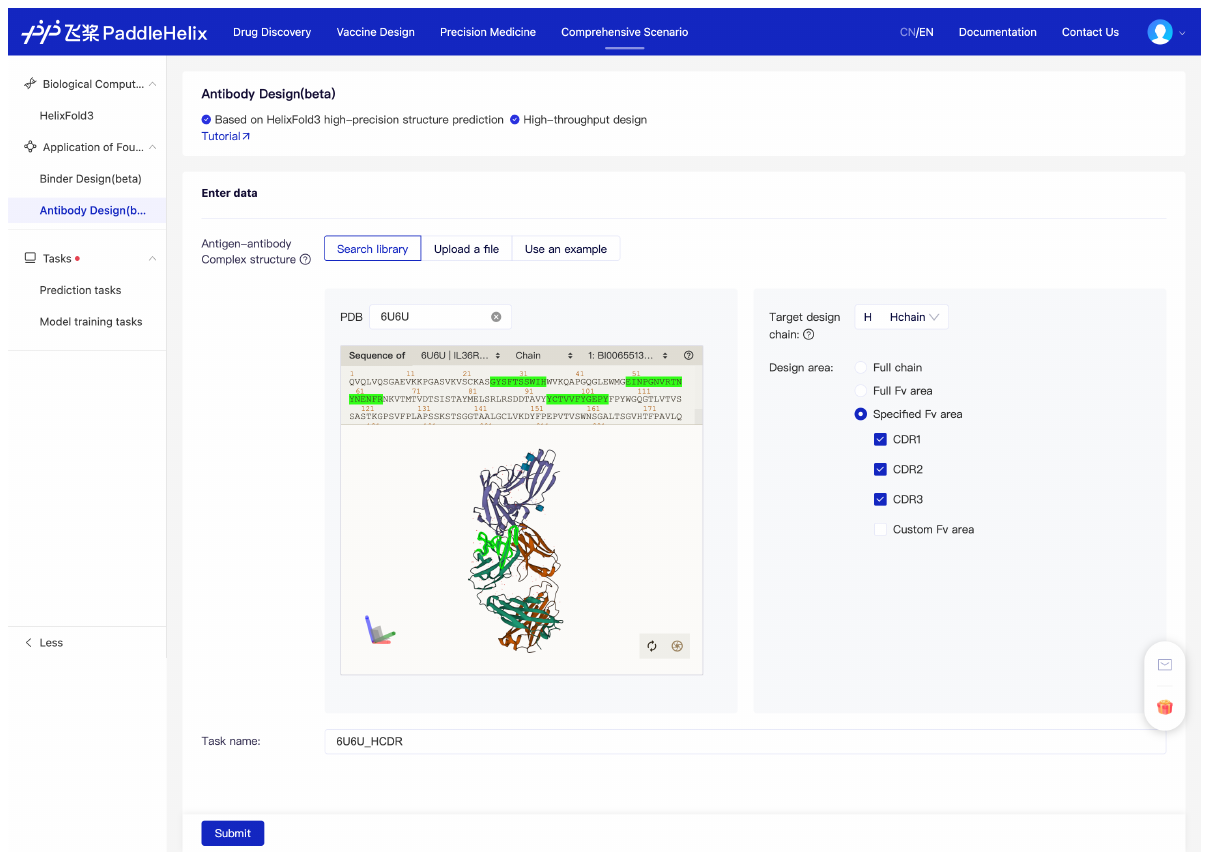}
    \caption{Example input interface of the HelixDesign-Antibody Server.
    }
    \label{fig:input_server}
\end{figure}

HelixDesign-Antibody (Figure ~\ref{fig:framework}) is a computational pipeline for structure-based antibody design that systematically generates and evaluates high-potential antibody candidates targeting specific antigens. The workflow supports the design of both full-length immunoglobulins (e.g., IgG) and nanobodies (single-domain antibodies), enabling flexible customization based on specific design requirements. 
As illustrated in the figure~\ref{fig:input_server}, 
The pipeline initiates by accepting a reference antibody-antigen complex structure as input, which can be either experimentally determined (e.g., from the Protein Data Bank, PDB\cite{bank1971protein}) or computationally predicted using structure prediction models such as HelixFold3, other folding algorithms, or docking tools for antigen-antibody interactions. Users define target design regions (e.g., CDR1–3 loops or full Fv domains) according to their objectives. Leveraging structural context, the pipeline then generates a diverse set of candidate sequences using structure-based sequence design models such as ESM-IF1\cite{hsu2022learning} and ProteinMPNN\cite{dauparas2022robust}. We currently utilize ESM-IF1 in HelixDesign-Antibody for candidate sequence design.

Existing design approaches typically require the generation of thousands of candidate candidate antibodies\cite{bennett2025atomically}. These candidates are subsequently undergo structure prediction for downstream filtering, resulting in considerable computational overhead. To address this challenge, HelixDesign-Antibody is optimized for high-throughput workflows in high-performance computing (HPC) environments, enabling the rapid screening of thousands of candidate antibodies within a single iteration. The diverse sequence design results are followed by high-throughput structure prediction of antibody-antigen complexes using production-grade HPC infrastructure. The predicted structures are rigorously evaluated through a multidimensional scoring system that integrates sequence-based, structure-based, and energy-based assessment methodologies, ultimately delivering a ranked list of promising candidates with optimized antigen recognition properties.

HelixDesign-Antibody is comprised of three integrated modules:
\begin{itemize}
    \item Structure-based Sequence Design: This module uses structural constraints to generate a wide range of candidate antibody sequences based on inverse folding algorithm (ESM-IF1).
    \item High-throughput Protein Folding (HelixFold3): This module predicts the 3D structures of the antigen–antibody complexes, assessing the structural reliability of each candidate.
    \item Multi-dimensional Interaction Analysis: This module evaluates and ranks the candidates based on multiple factors such as sequence-based scores, structure-based scores, and physicochemical scores.
\end{itemize}


\subsection{Structure-based Sequence Design}
In this module, the input consists of the reference antigen–antibody complex structure and user-defined designed regions (e.g., HCDR1). We adapt the ESM-IF1 \cite{hsu2022learning} inverse folding model, originally designed for the design of full-chain proteins. To generate antibody sequences based on these inputs, we modified the ESM-IF1 model to support the design of partial antibody sequence regions. Currently, the process generates 1,000 candidate antibody sequences per run. During sequence generation, the algorithm preserves the backbone conformation and the amino-acid sequences of non-designed regions, while proposing novel residues specifically within the designed regions to optimize structural compatibility. Each candidate sequence is assigned a fitness score, computed as the log-likelihood of residue-wise compatibility with the template structure. 
This structure-conditioned protein language model score reflects evolutionary fitness and is then used to select and prioritize the most promising antibody designs.

\subsection{High-throughput Protein Folding (HelixFold3)}
To evaluate the structural feasibility of designed antibody sequences, we use HelixFold3 \cite{liu2024technical}, a structure prediction model with accuracy comparable to, and in some cases exceeding, that of AlphaFold3 \cite{abramson2024accurate}, to generate 3D models of antibody–antigen complexes. The inputs to HelixFold3 include the amino acid sequence and structure of the target antigen, along with the amino acid sequences of the designed antibody candidates. Unlike AlphaFold3, the input structure of the target antigen can greatly enhance the accuracy of antigen-antibody complex conformation predictions and improve the reliability of conformation-based scoring. Additionally, other constraints, such as epitope information and residue contacts, can be incorporated to further improve HelixFold3’s conformation predictions. These inputs are then used to predict the resulting complex structures between each candidate and the antigen.

In addition to structural models, HelixFold3 outputs confidence metrics such as the interface predicted TM-score (ipTM) and inter-chain predicted aligned error (PAE). The utility of ipTM in identifying binding-competent antibodies was demonstrated in our previous work on HelixFold-Multimer \cite{fang2024helixfold}, and it continues to serve as a key indicator of interface quality. PAE provides complementary information by quantifying local uncertainty at the antibody–antigen interface.

Beyond general structure confidence, the predicted complexes are also used for epitope-level consistency checks. Epitope regions are inferred from the user-provided reference complex, and the pipeline evaluates whether the designed antibodies maintain contacts with these regions. This comparison helps identify candidates that preserve the intended binding mode, filtering out those that deviate despite appearing structurally valid.

To support large-scale evaluation, HelixDesign-Antibody automates the full prediction workflow, including sequence pairing and completion, input generation, model execution, and result aggregation. This enables efficient processing of multiple antibody–antigen complexes in parallel. By integrating seamlessly with the design module, the system ensures that only structurally viable candidates are advanced for downstream analysis, significantly reducing the experimental screening burden.

\subsection{Multi-dimensional Interaction Analysis}
Three major approaches have been developed for protein sequence evaluation. The first relies on protein language models trained on large-scale datasets of natural sequences, which infer evolutionary constraints and assign fitness scores that correlate with functional viability~\cite{notin2023proteingym, shanker2024unsupervised}. The second approach is based on structure-derived evaluation of interface plausibility, using metrics such as the predicted interface TM-score (ipTM). This strategy has proven critical for improving the success rates of de novo protein design~\cite{zambaldi2024novo, watson2023novo}. The third approach employs energy-based methods, such as Rosetta~\cite{barlow2018flex} and FoldX~\cite{schymkowitz2005foldx}, to estimate the favorability of physicochemical binding properties in candidate structures.

HelixDesign-Antibody integrates these multi-dimensional evaluation strategies by combining three complementary metrics: sequence-based fitness scores, structure-based interface metrics, and physicochemical scores. When epitope information for the antigen is available, predicted epitope contact patterns are also incorporated to further refine candidate selection.

Candidates are systematically ranked and filtered based on consistency across evaluation metrics and the specific priorities of the design objectives. Manual validation of epitope contacts remains essential during the final selection process. Designs that exhibit strong interactions with conserved or functionally critical epitope residues are prioritized, as such contacts are often indispensable for effective antigen recognition. This hybrid computational and manual workflow ensures that selected candidates meet both biophysical standards and structural-functional relevance.

\begin{figure}[htbp]
    \centering
        \includegraphics[width=\linewidth]{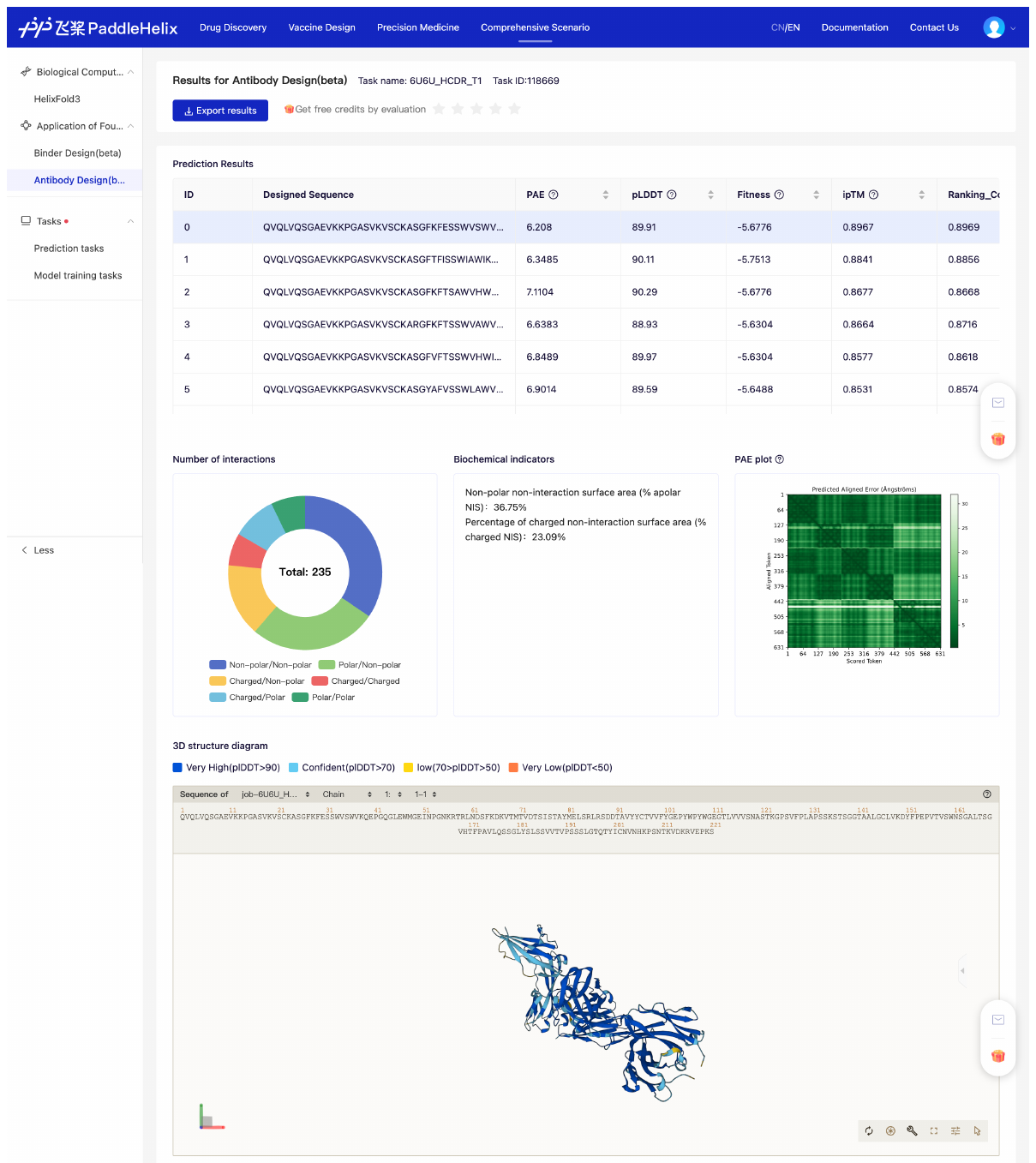}        
    \caption{Example output display page of the HelixDesign-Antibody Server.
    }
    \label{fig:output_server}
\end{figure}

To facilitate intuitive interpretation of the prediction results, we provide visualizations of key evaluation metrics, as shown in Figure~\ref{fig:output_server}. Such comprehensive visualization enables users to quickly assess the quality and plausibility of each antibody design.

\section{Results}

\begin{figure}[ht]
    \centering

    \begin{minipage}[c]{\textwidth}
        \centering
        \begin{subfigure}[t]{0.31\textwidth}
            \centering
            \includegraphics[height=4.1cm]{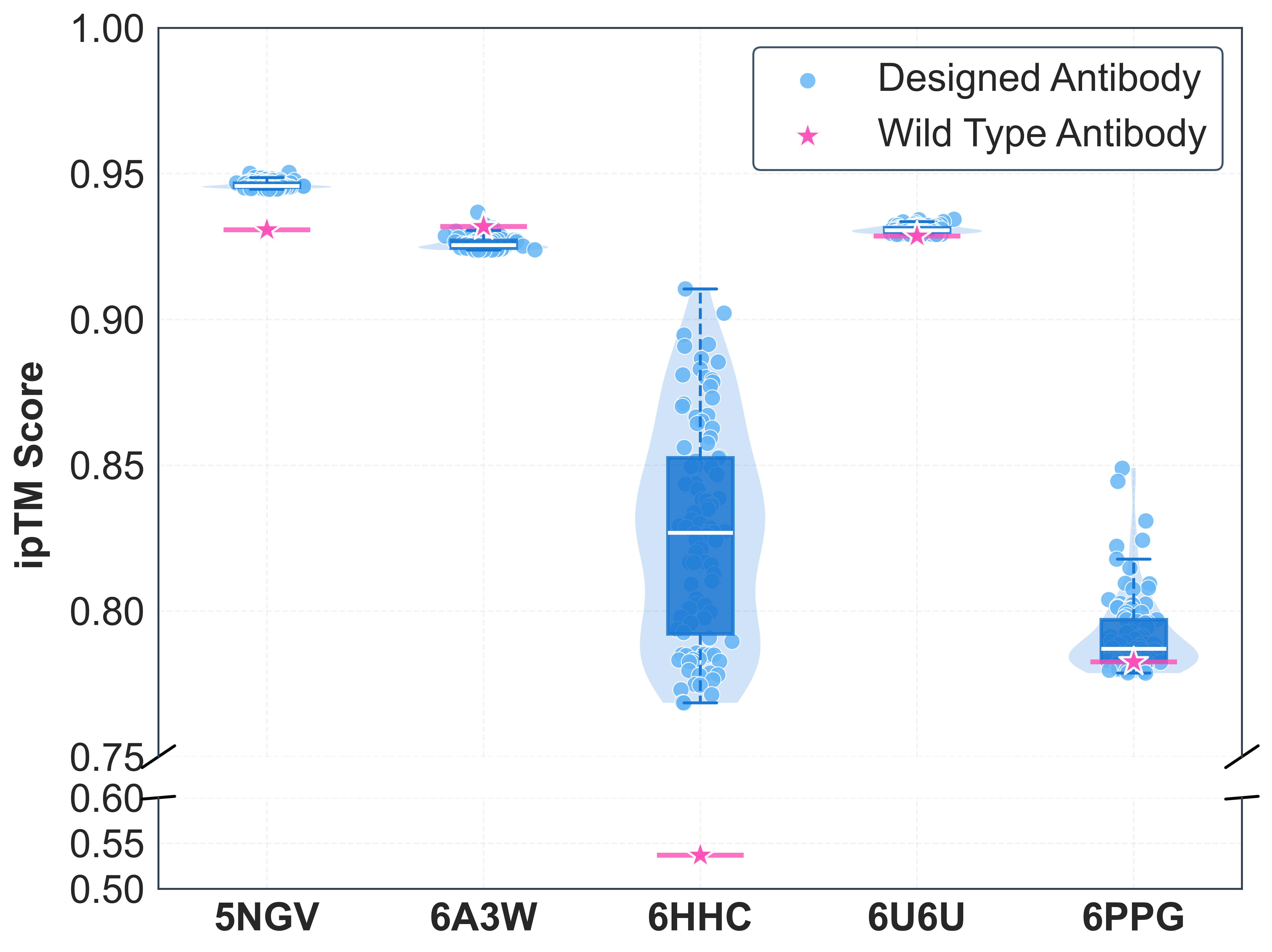}
            \caption{Interface predicted TM-score}
            \label{fig:iptm}
        \end{subfigure}
        \hfill
        \begin{subfigure}[t]{0.31\textwidth}
            \centering
            \includegraphics[height=4.1cm]{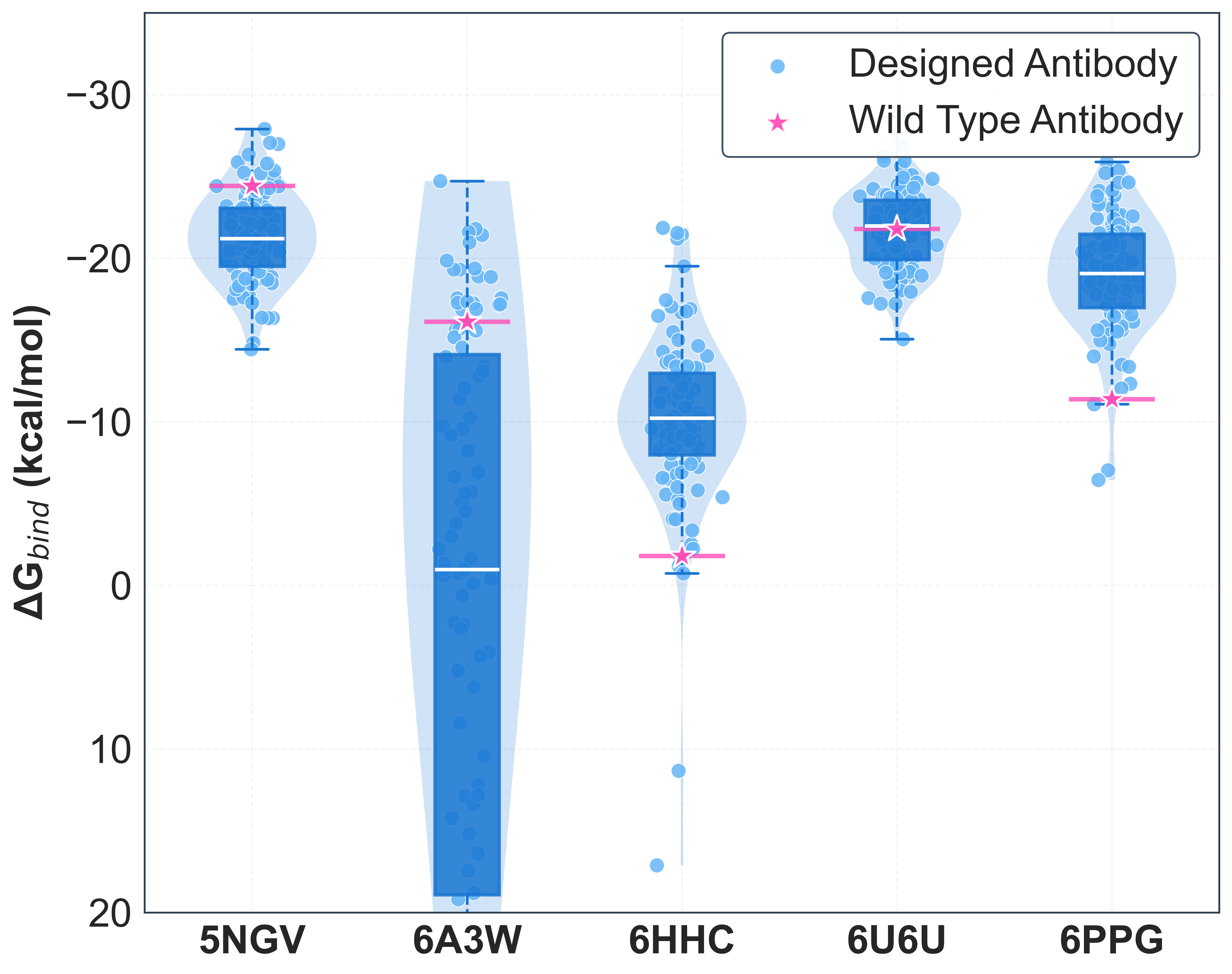}
            \caption{Predicted binding free energy}
            \label{fig:dgfoldx}
        \end{subfigure}
        \hfill
        \begin{subfigure}[t]{0.31\textwidth}
            \centering
            \includegraphics[height=4.1cm]{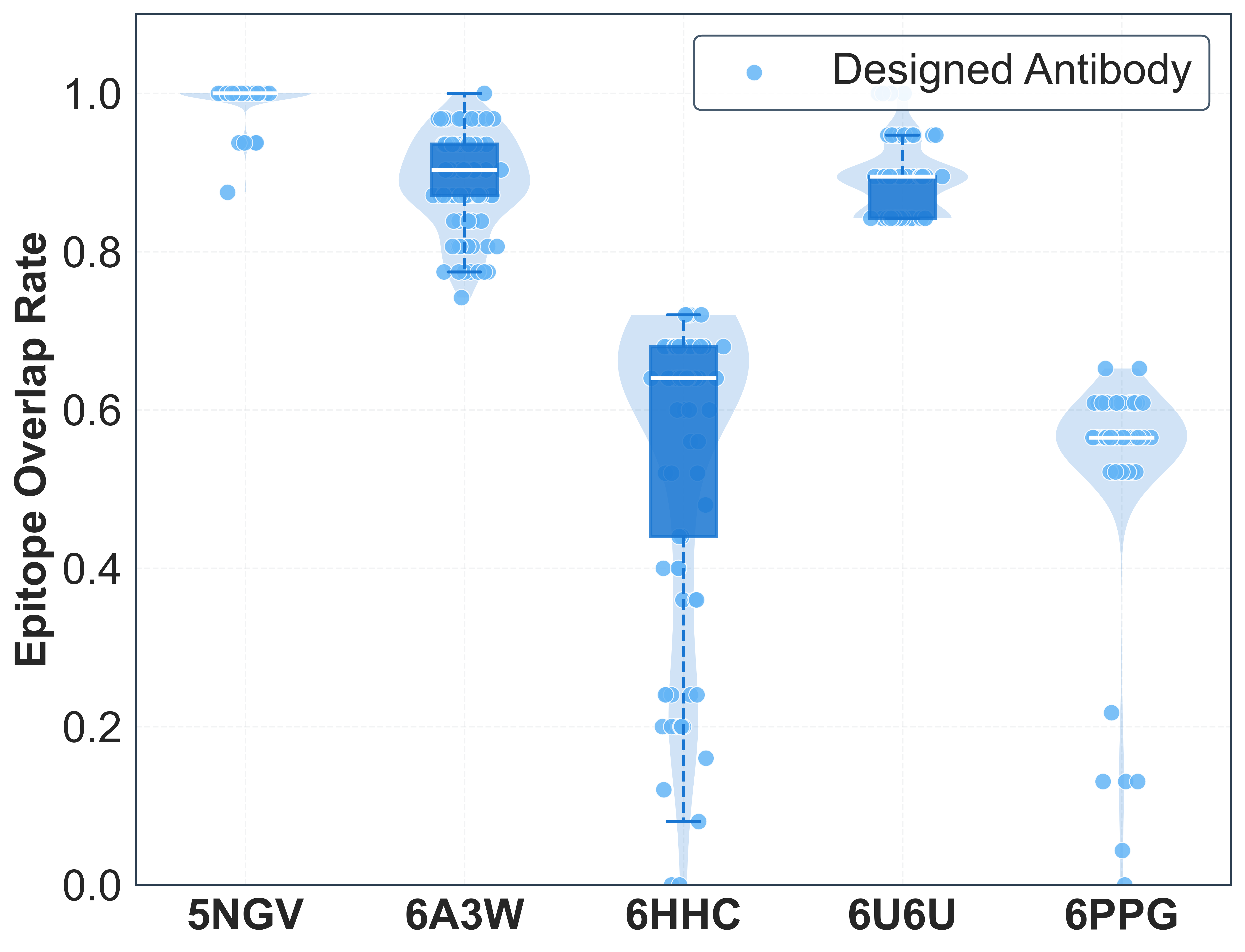}
            \caption{Epitope Overlap}
            \label{fig:epitope_overlap}
        \end{subfigure}
    \end{minipage}

    \vspace{1em}

    \begin{minipage}[c]{\textwidth}
        \centering
        \begin{subfigure}[t]{0.675\textwidth}
            \centering
            \includegraphics[width=\linewidth]{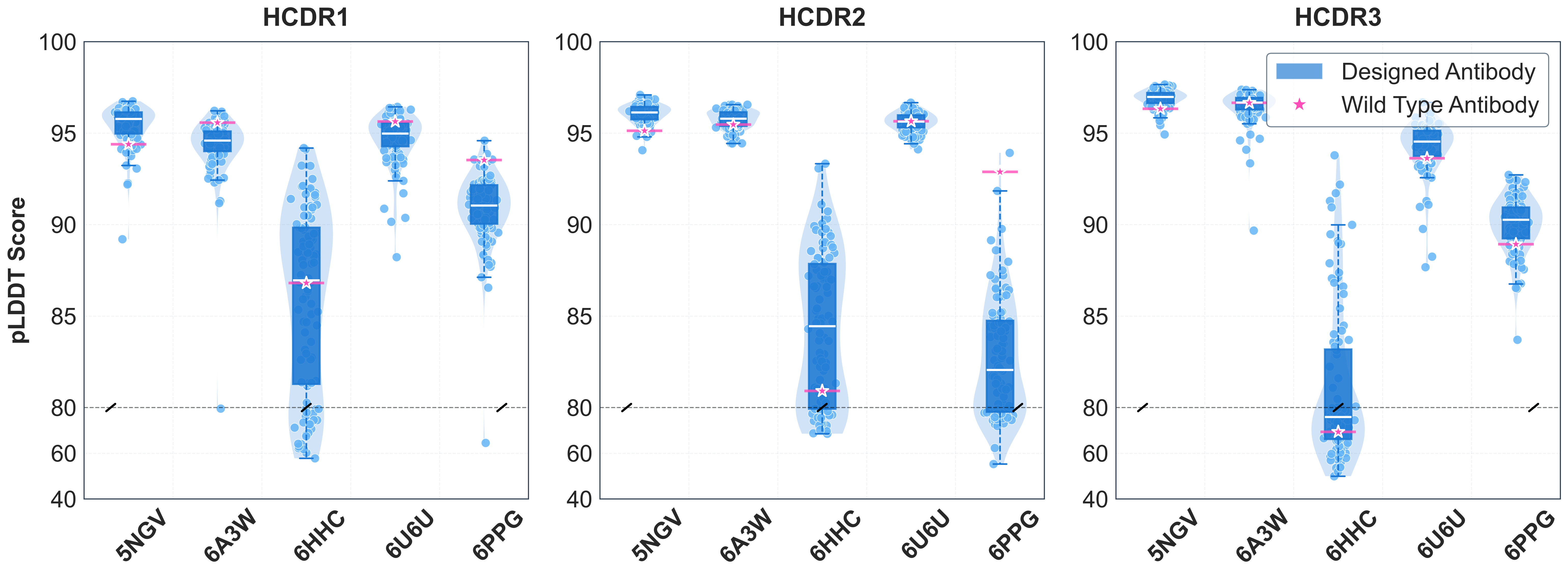}
            \caption{CDR pLDDT}
            \label{fig:plddt}
        \end{subfigure}
        \hfill
        \hfill
        \hfill
        \begin{subfigure}[t]{0.315\textwidth}
            \centering
            \includegraphics[width=\linewidth]{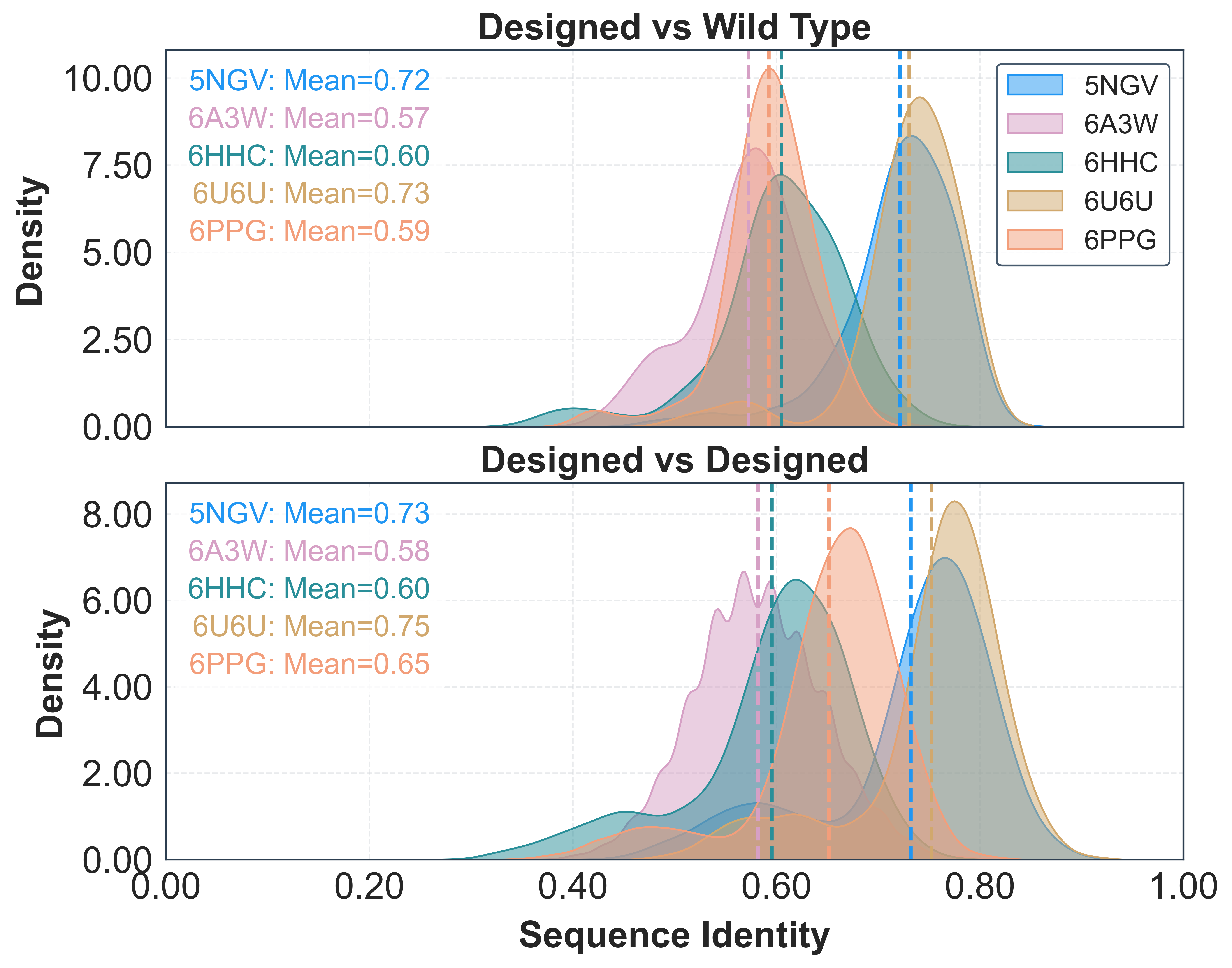}
            \caption{Diversity and Novelty}
            \label{fig:diveristy}
        \end{subfigure}
    \end{minipage}

    \caption{Comparison and evaluation of designed antibody sequences across multiple metrics.  The top row shows quantitative metrics of binding quality across designs, including (a) interface predicted TM‑scores (ipTM), (b) predicted binding free energies ($\Delta G$), and (c) epitope overlap with wild‑type sequences. The bottom row evaluates structural quality and sequence diversity, highlighting (d) CDR pLDDT as a measure of local structural confidence, and (e) the distribution of designed sequences relative to wild‑type, capturing both novelty and diversity. Together, these metrics underscore the benefits of large‑scale design in achieving higher‑quality, structurally confident, and diverse antibody candidates across multiple antigens.}
    \label{fig:design_summary}
\end{figure}

\subsection{Overall Performance of Antibody Design}
To evaluate the performance of HelixDesign-Antibody, we conducted design experiments on five antigens with available co-crystal structures and their corresponding reference antibodies: ACVR2B (PDB: 5NGV), TNFRSF9 (PDB: 6A3W), FXI (PDB: 6HHC), IL-36R (PDB: 6U6U), and IL17A (PDB: 6PPG). For each case, we used the backbone structure of the reference antibody as a template and optimized only the complementarity-determining regions (CDRs) of the heavy chain. The light chain and the framework regions—i.e., the non-CDR residues within the variable domains—were kept fixed to maintain structural consistency with the original antibody.

We employed two complementary metrics—structural confidence and binding free energy to evaluate the quality of designed antibody–antigen complexes. Structural confidence was assessed using the interface predicted Template Modeling (ipTM) score and pLDDT scores of the CDR regions, computed by HelixFold3 \cite{liu2024technical}, a structure prediction model capable of accurately modeling antibody–antigen interfaces. Binding free energy ($\Delta G$), estimated by FoldX \cite{schymkowitz2005foldx}, provides a physically grounded measure of interaction strength. Together, these metrics capture both the geometric plausibility and energetic viability of each design, which are critical for identifying candidates with high potential for functional binding. This dual-metric framework is well suited for high-throughput in silico antibody design, where balancing structural accuracy with predicted affinity is essential for effective prioritization.

A comparative analysis of the designed and wild-type antibodies is presented in Figure~\ref{fig:iptm}, Figure~\ref{fig:plddt}, and Figure~\ref{fig:dgfoldx}. Across all five antigens, the designed antibodies achieve ipTM scores that are comparable to those of the corresponding wild-type antibodies. Notably, for four antigens, the ipTM scores of the designed antibodies exceed 0.7, suggesting a high level of structural confidence in the predicted antibody–antigen interfaces according to HelixFold3. The pLDDT scores of the CDR regions in the designed antibodies are generally above 80, indicating well-defined and stable conformations for these critical binding loops. In addition, the designed antibodies exhibit more favorable (i.e., lower) binding free energy ($\Delta G$) values than the wild-type antibodies in all five antigens, indicating potential improvements in predicted binding affinity. These results suggest that HelixDesign-Antibody can generate candidates with structurally plausible interfaces and favorable energetic profiles, supporting its utility in early-stage, in silico antibody design workflows.

To investigate whether the designed antibodies are likely to retain the biological functions of their wild-type counterparts, we analyzed their binding interfaces with the target antigens. The wild-type antibody–antigen interfaces were obtained from experimentally determined co-crystal structures, while those of the designed antibodies were derived from structures predicted by HelixFold3. We adopted a broad definition of epitope, identifying antigen-contacting residues as those located within 8Å of any antibody residue. Based on this definition, we computed the proportion of overlapping epitope residues between the wild-type and designed antibodies as a proxy for functional similarity at the interface. Across the five antigen targets, the designed antibodies exhibited moderate to high levels of epitope overlap with their wild-type counterparts, suggesting that HelixDesign-Antibody can generate candidates that potentially preserve key antigen recognition sites and thus retain biological activity.

\subsection{Diversity and Novelty of the Designed Antibodies}
In Figure~\ref{fig:diveristy}, we compare the CDR sequence identities of the designed antibodies both to their respective wild‑type templates and within the designed sets across five antigens. The top panel shows that the designed antibodies retain a moderate level of similarity to their wild‑type templates, with average CDR identities ranging from approximately 57\% (6A3W) to 73\% (5NGV, 6U6U). This balance ensures that the designs maintain key structural characteristics of the original scaffolds while introducing substantial sequence variability. The bottom panel highlights the diversity within the designed sets, with average CDR identities generally clustered between 58\% and 75\%, indicating a broad exploration of sequence space. Together, these results demonstrate that HelixDesign‑Antibody effectively balances structural conservation and sequence diversity in the CDRs, providing a strong foundation for identifying high‑quality, target‑specific candidates.

\begin{figure}[htbp]
    \centering

    \text{Target: TNFRSF9 (6A3W)}\\[0.5em]
    \begin{subfigure}{0.23\textwidth}
        \centering
        \includegraphics[width=\linewidth]{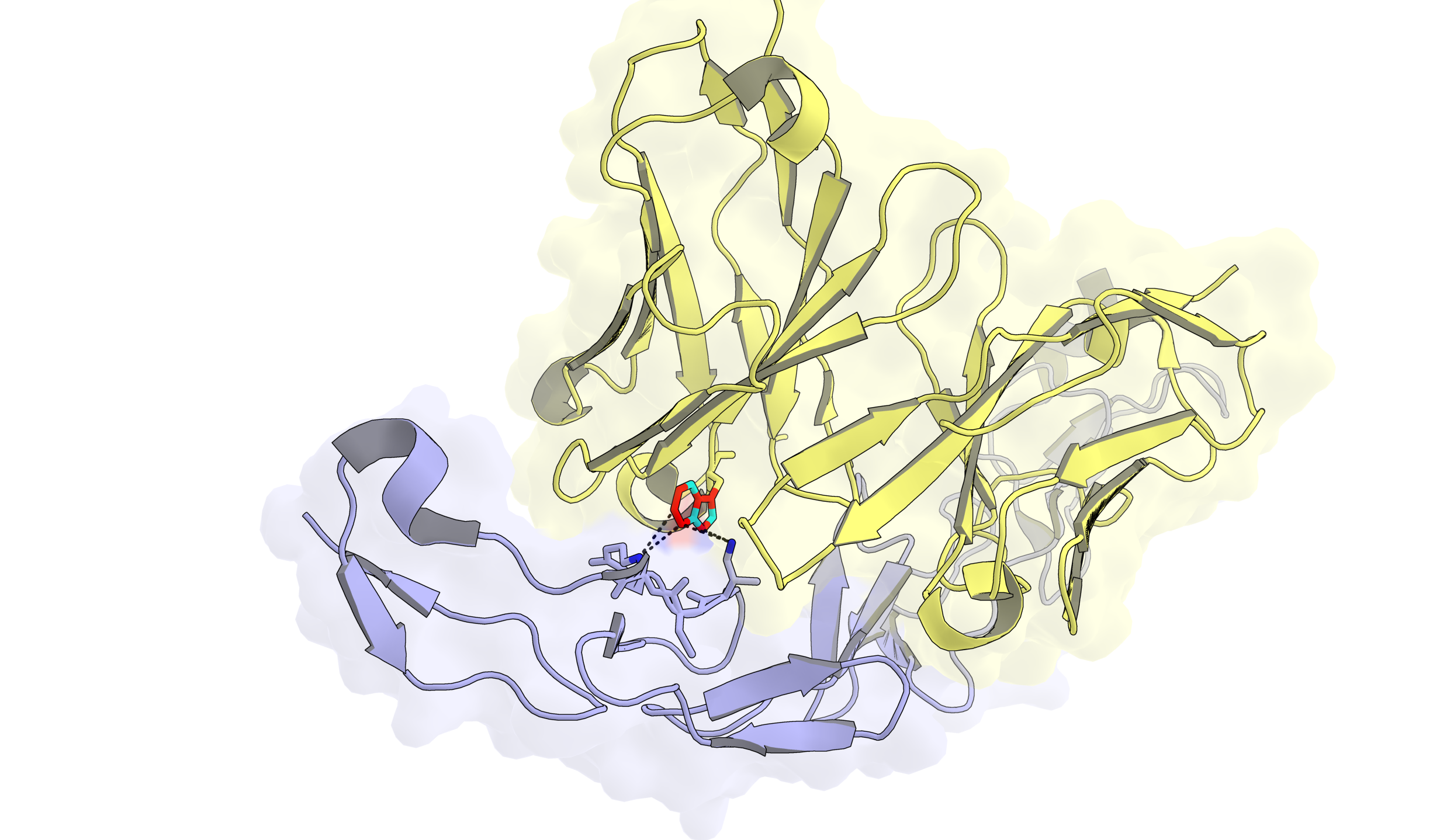}
        \caption{Wild-type structure}
        \label{fig:6A3W_wild_global}
    \end{subfigure}
    \hfill
    \begin{subfigure}{0.25\textwidth}
        \centering
        \includegraphics[width=\linewidth]{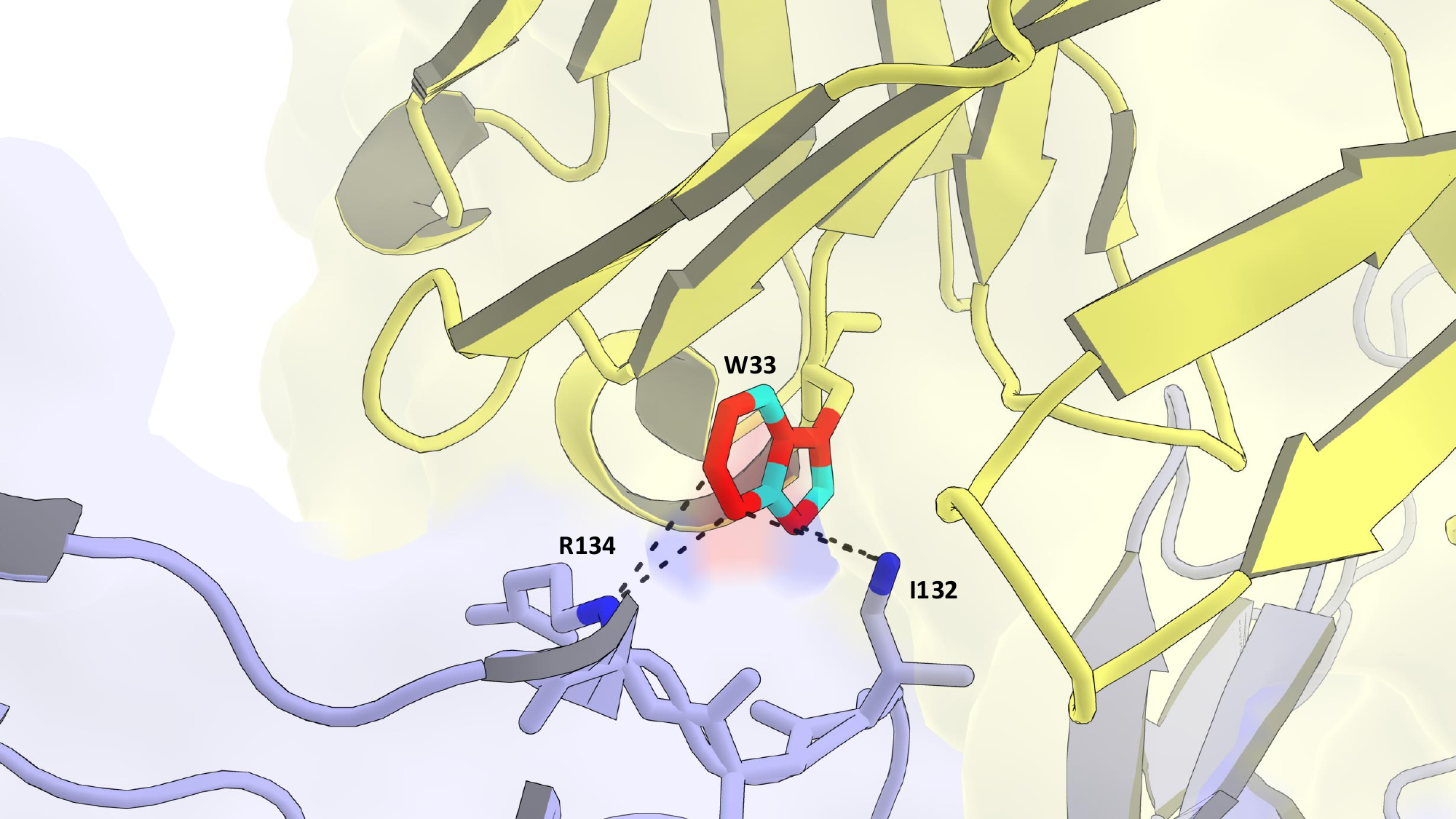}
        \caption{Wild-type interaction}
        \label{fig:6A3W_wild_local}
    \end{subfigure}
    \hfill
    \begin{subfigure}{0.23\textwidth}
        \centering
        \includegraphics[width=\linewidth]{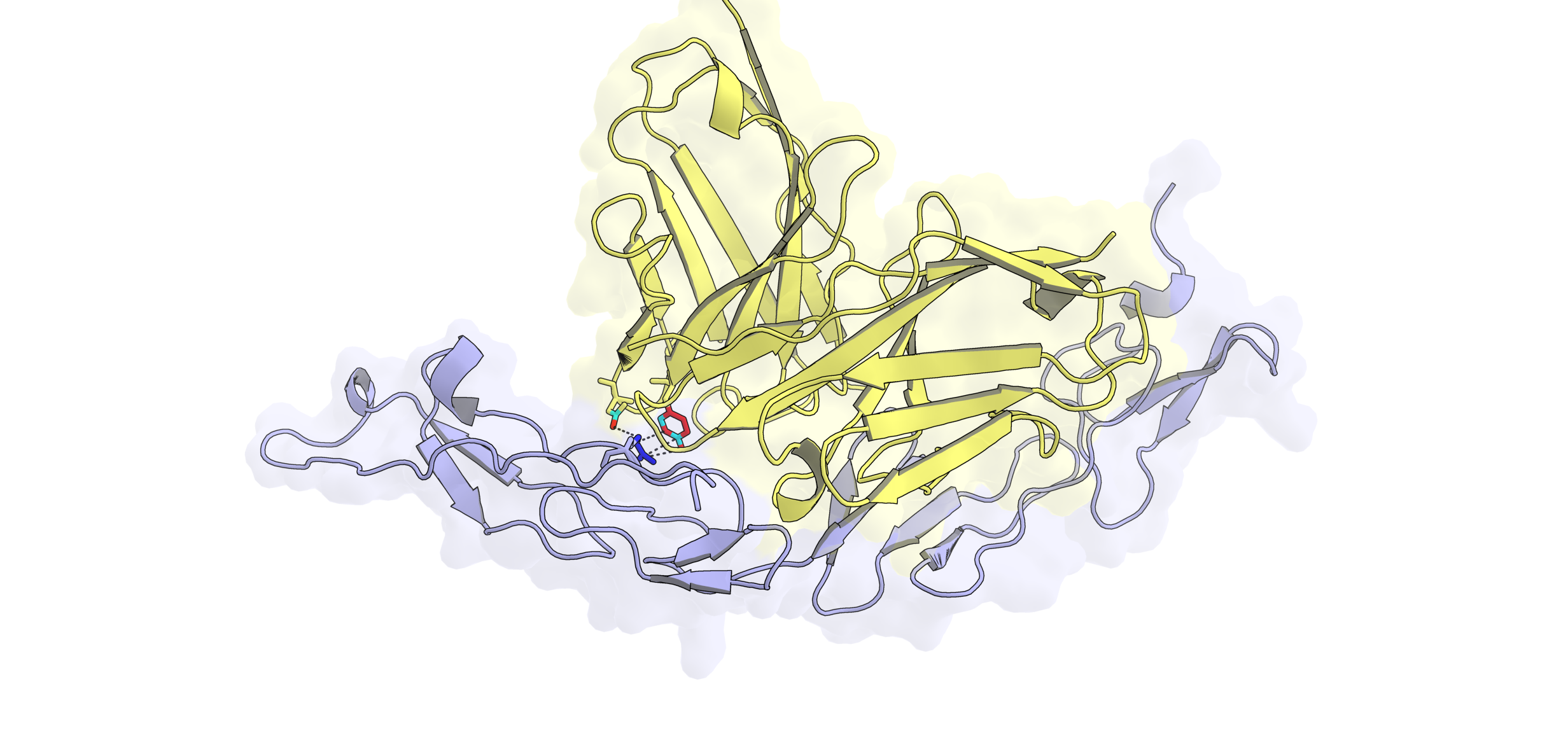}
        \caption{Designed structure}
        \label{fig:6A3W_design_global}
    \end{subfigure}
    \hfill
    \begin{subfigure}{0.25\textwidth}
        \centering
        \includegraphics[width=\linewidth]{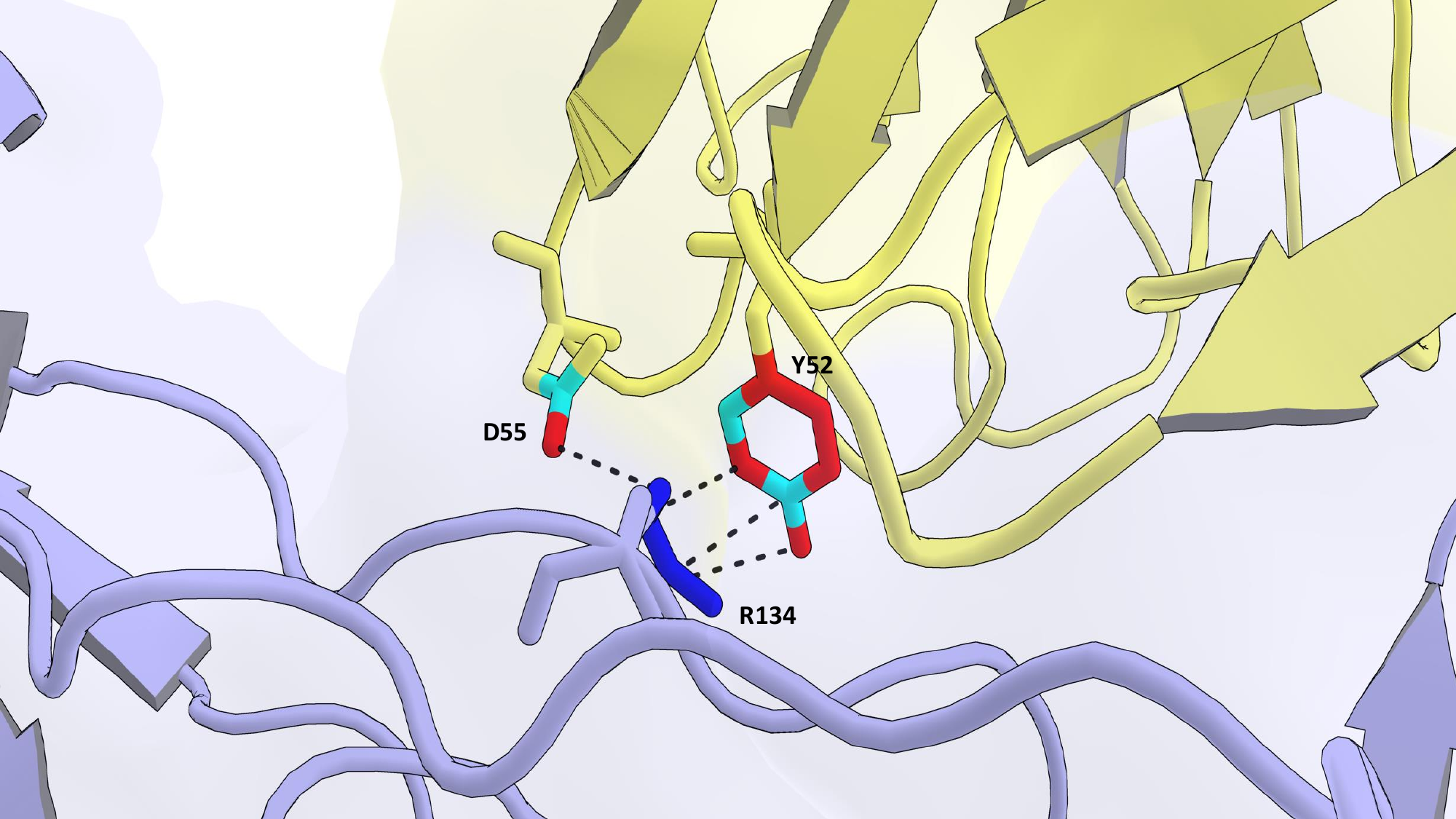}
        \caption{Designed interaction}
        \label{fig:6A3W_design_local}
    \end{subfigure}
    \text{Target: IL-36R (6U6U)}\\[0.5em]
    \begin{subfigure}{0.23\textwidth}
        \centering
        \includegraphics[width=\linewidth]{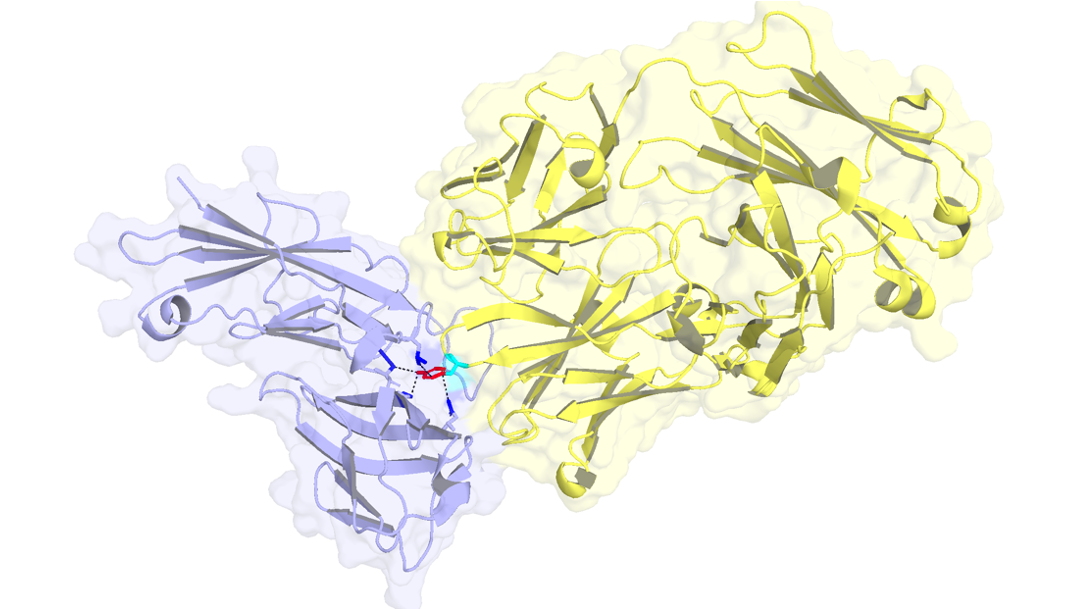}
        \caption{Wild-type structure}
        \label{fig:6u6u_wild_global}
    \end{subfigure}
    \hfill
    \begin{subfigure}{0.25\textwidth}
        \centering
        \includegraphics[width=\linewidth]{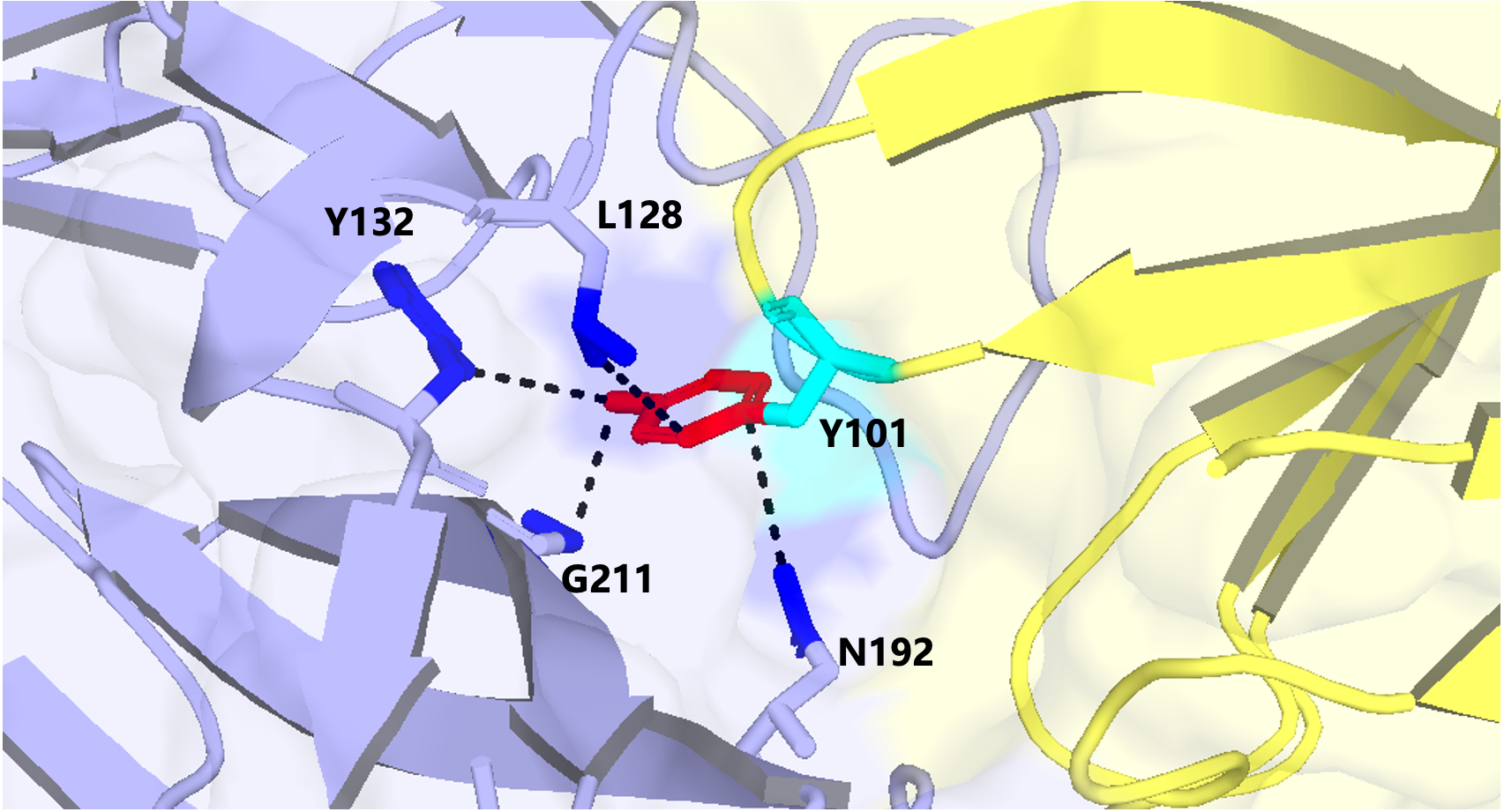}
        \caption{Wild-type interaction}
        \label{fig:6u6u_wild_local}
    \end{subfigure}
    \hfill
    \begin{subfigure}{0.23\textwidth}
        \centering
        \includegraphics[width=\linewidth]{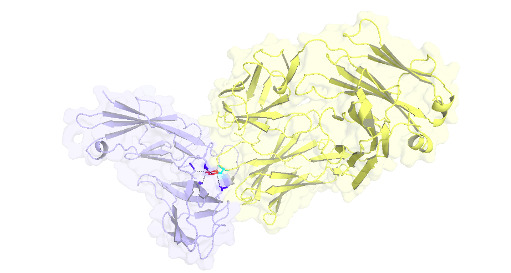}
        \caption{Designed structure}
        \label{fig:6u6u_design_global}
    \end{subfigure}
    \hfill
    \begin{subfigure}{0.25\textwidth}
        \centering
        \includegraphics[width=\linewidth]{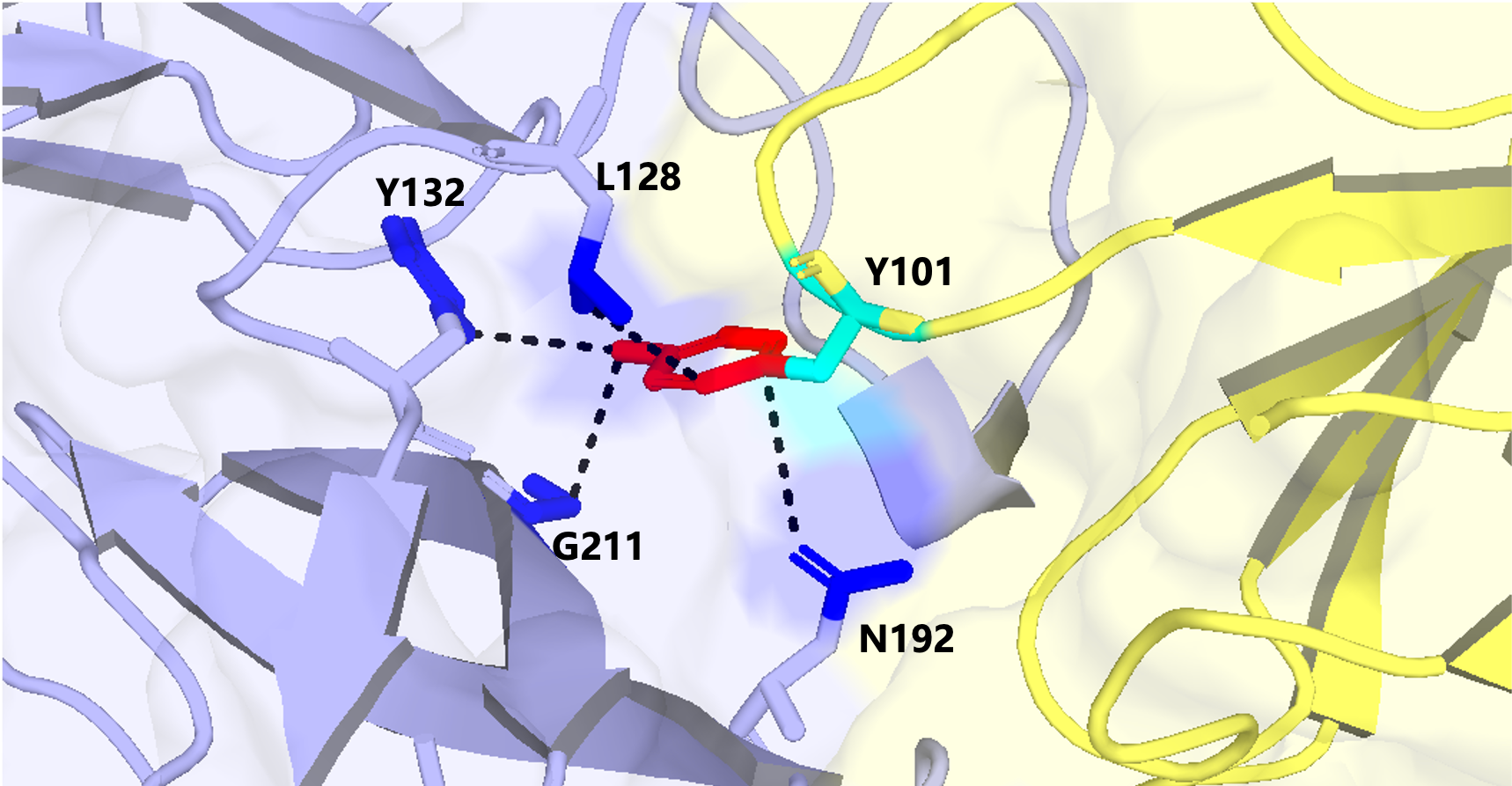}
        \caption{Designed interaction}
        \label{fig:6u6u_design_local}
    \end{subfigure}

    \caption{
Structural comparison of wild-type and designed antibodies for two targets. 
\textbf{Top:} Targeting TNFRSF9 (PDB: 6A3W). 
\textbf{Bottom:} Targeting IL-36R (PDB: 6U6U). 
For each target, the left two panels show the wild-type antibody structure and its binding interface, and the right two panels show the designed version. 
Antigen interaction atoms are shown in blue, while antibody atoms are colored red and cyan. 
Zoomed-in panels highlight key interface residues, with dashed lines indicating hydrogen bonds. Residue indices are labeled numerically and amino acids by one-letter codes.
    }
    \label{fig:design_case}
\end{figure}

We visualized the antibody–antigen complexes for two targets, TNFRSF9 (PDB ID: 6A3W) and IL-36R (PDB ID: 6U6U), and compared the binding modes of the wild-type and designed antibodies (Figure~\ref{fig:design_case}).
For 6A3W (Figure~\ref{fig:design_case}a-d), in the wild-type structure, the antibody utilizes W33 to engage the epitope via aromatic interactions. In the designed case, this position is replaced by Y52, preserving an aromatic side chain within the binding interface while slightly altering its orientation and local packing. This suggests that HelixDesign–Antibody can maintain the key physicochemical character of the binding site despite differences in side chain identity. Additionally, polar residues such as R134 and D55 are positioned proximally to the antigen surface, implying the potential for additional polar or hydrogen-bonding interactions. 

For 6U6U (Figure~\ref{fig:design_case}e–h), the wild-type structure reveals a compact interface network involving residues L128, Y132, N192, and G211. L128 and Y132, located within the D1–D2 linker of IL-36R, contribute to hydrophobic and hydrogen bonding interactions that anchor the antibody within the interdomain cleft~\cite{larson2020x}. N192 and G211, situated in the D2 domain, provide additional polar contacts that further stabilize the interface.
Despite sequence variations in other CDR regions, the designed antibody preserves the essential interaction mediated by Y101 within the CDRH3 loop, a residue known to be essential in the parental antibody BI 655130. This side chain inserts into the interdomain pocket between D1 and D2, preserving shape complementarity and hydrogen-bonding interactions with residues such as G211 and N192. 

Together, these observations highlight the ability of the design method to recapitulate critical epitope recognition motifs, while introducing subtle variations in binding geometry and residue composition.

\subsection{Computational Scale Enables Antibody Optimization}

\begin{figure}[H]
    \centering

    \begin{minipage}[c]{\textwidth}
        \centering
        \begin{subfigure}[t]{0.45\textwidth}
            \centering
            \includegraphics[width=\linewidth]{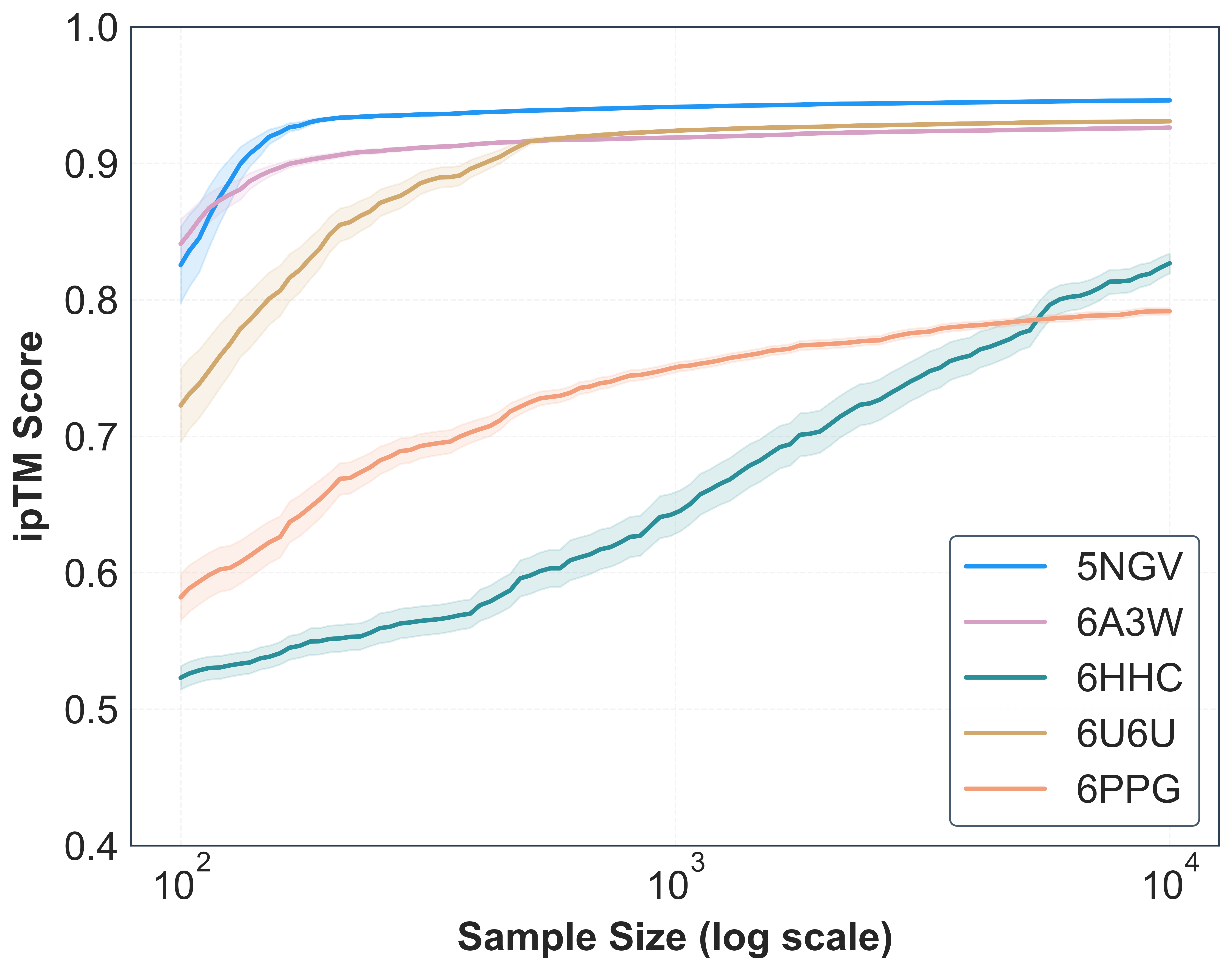}
            \caption{Improvement in ipTM as a function of sampling size}
            \label{fig:scaling_iptm}
        \end{subfigure}
        \hfill
        \begin{subfigure}[t]{0.45\textwidth}
            \centering
            \includegraphics[width=\linewidth]{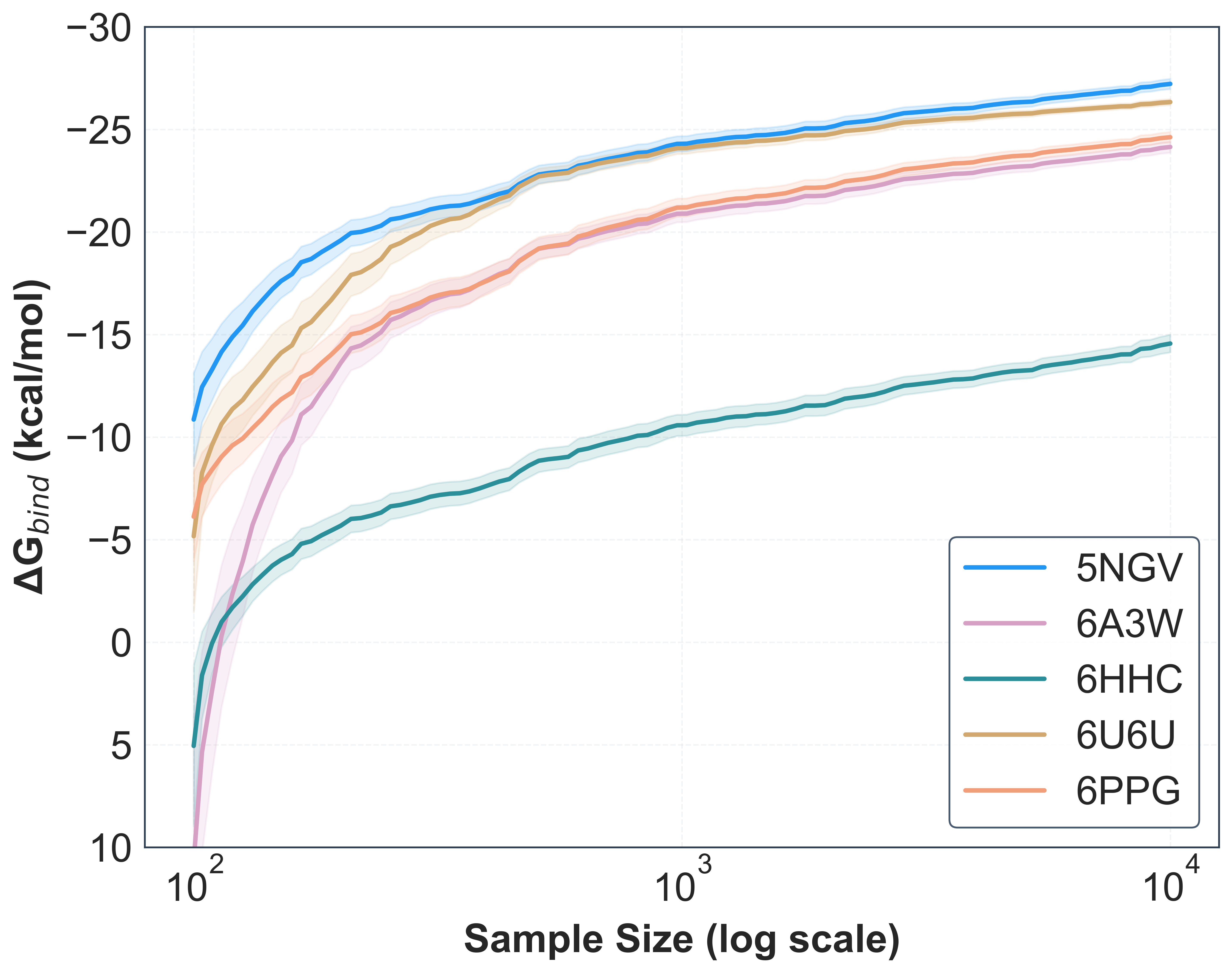}
            \caption{Improvement in binding free energy as a function of sampling size}
            \label{fig:scaling_dg}
        \end{subfigure}
    \end{minipage}
    \caption{Computational scale enables antibody optimization. Performance improvement as a function of sampling size across five different antibody-antigen systems (5NGV, 6A3W, 6HHC, 6U6U, and 6PPG). (a) Improvement in ipTM (interface predicted Template Modeling) score showing the quality of protein-protein interface prediction. (b) Improvement in binding free energy expressed in kcal/mol.}
    \label{fig:scaling}
\end{figure}

In HelixDesign‑Antibody, we observe a clear quantitative relationship between the number of designed sequences and the quality of predicted antibody–antigen complexes across five targets (Figure~\ref{fig:scaling}). For each target, the solid line represents the mean ipTM score, while the shaded region denotes the 95\% confidence interval. As the sampling size increases, both the average predicted binding free energies and the ipTM scores of the top‑ranked 100 designs consistently improve across all targets. This trend highlights the pivotal role of large‑scale design: by extensively exploring the antibody sequence space, we substantially raise the likelihood of identifying candidates with favorable biophysical and structural characteristics.

Importantly, these results echo the motivation stated in the introduction: HelixDesign‑Antibody leverages Baidu AI Cloud’s HPC platform to enable high‑throughput generation and evaluation of antibody candidates. The ability to efficiently scale design and screening not only improves the quality of the top‑ranked candidates but also provides a robust foundation for antibody engineering pipelines that can be applied to a broad range of targets.

\subsection{Binding Discrimination Capability}
HelixFold3 generates structure-based confidence scores, including the interface predicted Template Modeling (ipTM) score, which quantifies the quality of predicted intermolecular interfaces. To evaluate its utility in binding discrimination, we focused on the task of predicting relative binding strength among antibody–antigen complexes.

\begin{figure}[H]
    \centering
    \includegraphics[width=0.7\linewidth]{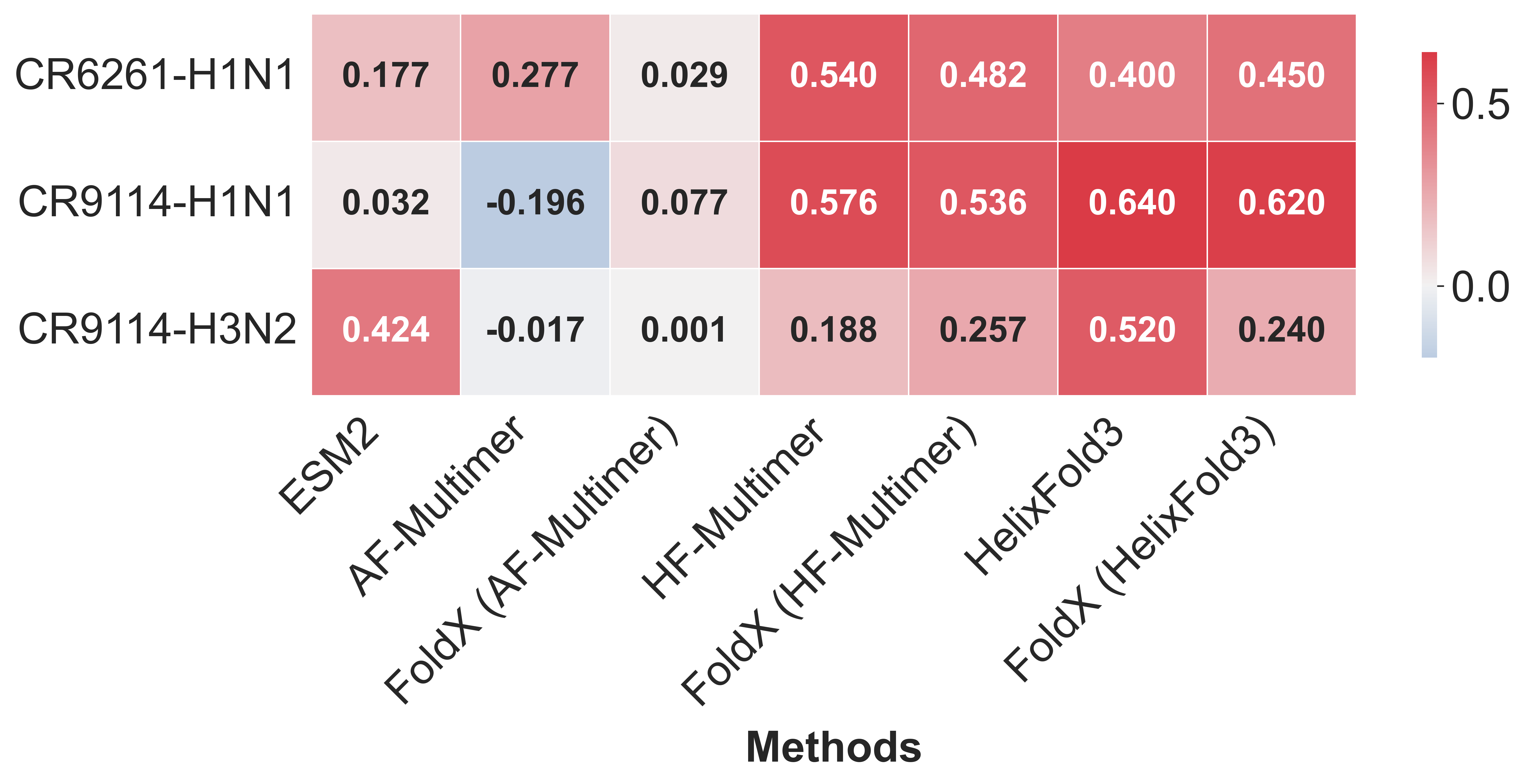}
    \caption{Binding discrimination capability of different scoring methods across datasets. Heatmap showing Pearson correlation coefficients (PCC) between predicted scores from various methods and experimental binding affinities. The analysis was performed on three protein-ligand datasets: CR6261-H1N1, CR9114-H1N1, and CR9114-H3N2. Color intensity represents correlation strength, with red indicating positive correlations and blue indicating negative correlations.}
    \label{fig:binding_discrimination}

\end{figure}

Specifically, we benchmarked HelixFold3 on two influenza broadly neutralizing antibodies, CR9114~\cite{beukenhorst2022influenza} and CR6261~\cite{ekiert2009antibody} , which target the conserved stem epitope of hemagglutinin (HA). These antibodies have experimentally determined dissociation constants ($K_D$) across different viral subtypes~\cite{phillips2021binding}.
For CR6261, we used all available affinity data against H1N1 with approximately 2000 samples. For CR9114, to reduce computational complexity, we sampled 2000 samples for each target(H1N1 and H3N2) based on the distribution of the affinity data. For the H1N1 affinity data, we randomly sampled 2000 samples that replicate the overall distribution of affinity across all data. For H3N2, due to its highly imbalanced affinity data distribution, where high-affinity samples are less frequent, we applied a 5:1 weighting to more frequently sample high-affinity examples, ensuring some degree of sample balance. Ultimately, we obtained about 2000 samples for each type of affinity data for CR6261 (H1N1) and CR9114 (H1N1, H3N2).
The structure predicted with HelixFold3 used a reference antigen structure to improve the accuracy.
Ultimately, HelixFold3 achieved Pearson correlation coefficients of 0.40, 0.64, and 0.52 for CR6261–H1N1, CR9114–H1N1, and CR9114–H3N2, respectively—substantially outperforming AlphaFold-Multimer~\cite{evans2021protein} and ESM-2~\cite{lin2022language} (Figure~\ref{fig:binding_discrimination}). Binding free energies computed by FoldX based on the structures predicted by HelixFold3 also demonstrated improved performance compared to those based on AlphaFold-Multimer predictions.
These results demonstrate that the ipTM score effectively captures binding-relevant structural differences and can distinguish between stronger and weaker binders. Even in the context of subtle sequence variants, HelixFold3 maintains robust sensitivity to affinity changes, highlighting ipTM as a practical and versatile metric for antibody binding evaluation.

\section{Conclusion and Future Works}
We have developed HelixDesign-Antibody, a scalable platform that leverages HelixFold3 for efficient antibody design, binding evaluation, and stability assessment. The platform has demonstrated its ability to generate diverse, high-quality antibodies across multiple antigens, highlighting the effectiveness of large-scale sequence exploration in identifying optimal binders.

Looking ahead, we aim to expand the platform's capabilities by incorporating de novo antibody design, further improving structural prediction accuracy, and refining design modules to support more complex design scenarios. Additionally, we plan to integrate a wider range of scoring functions to evaluate not only binding affinity and stability but also manufacturability and other critical properties, making the platform a more comprehensive tool for antibody optimization.

We welcome both academic and industry scientists to explore and trial the platform, and encourage feedback to help us address challenges and further improve the system. We are also open to various types of collaborations, and look forward to working together with partners to advance the field of antibody design. For inquiries and collaboration opportunities, please contact us at \href{mailto:baidubio_cooperate@baidu.com}{baidubio\_cooperate@baidu.com}.

\clearpage

\clearpage

\bibliographystyle{unsrt}  
\bibliography{references}

\begin{thebibliography}{10}

\bibitem{evans2021protein}
Richard Evans, Michael O’Neill, Alexander Pritzel, Natasha Antropova, Andrew Senior, Tim Green, Augustin {\v{Z}}{\'\i}dek, Russ Bates, Sam Blackwell, Jason Yim, et~al.
\newblock Protein complex prediction with alphafold-multimer.
\newblock {\em biorxiv}, pages 2021--10, 2021.

\bibitem{abramson2024accurate}
Josh Abramson, Jonas Adler, Jack Dunger, Richard Evans, Tim Green, Alexander Pritzel, Olaf Ronneberger, Lindsay Willmore, Andrew~J Ballard, Joshua Bambrick, et~al.
\newblock Accurate structure prediction of biomolecular interactions with alphafold 3.
\newblock {\em Nature}, 630(8016):493--500, 2024.

\bibitem{gao2024precise}
Jie Gao, Jing Hu, Lihang Liu, Yang Xue, Kunrui Zhu, Xiaonan Zhang, and Xiaomin Fang.
\newblock Precise antigen-antibody structure predictions enhance antibody development with helixfold-multimer.
\newblock {\em arXiv preprint arXiv:2412.09826}, 2024.

\bibitem{liu2024technical}
Lihang Liu, Shanzhuo Zhang, Yang Xue, Xianbin Ye, Kunrui Zhu, Yuxin Li, Yang Liu, Jie Gao, Wenlai Zhao, Hongkun Yu, et~al.
\newblock Technical report of helixfold3 for biomolecular structure prediction.
\newblock {\em arXiv preprint arXiv:2408.16975}, 2024.

\bibitem{hsu2022learning}
Chloe Hsu, Robert Verkuil, Jason Liu, Zeming Lin, Brian Hie, Tom Sercu, Adam Lerer, and Alexander Rives.
\newblock Learning inverse folding from millions of predicted structures.
\newblock {\em ICML}, 2022.

\bibitem{dauparas2022robust}
Justas Dauparas, Ivan Anishchenko, Nathaniel Bennett, Hua Bai, Robert~J Ragotte, Lukas~F Milles, Basile~IM Wicky, Alexis Courbet, Rob~J de~Haas, Neville Bethel, et~al.
\newblock Robust deep learning--based protein sequence design using proteinmpnn.
\newblock {\em Science}, 378(6615):49--56, 2022.

\bibitem{schymkowitz2005foldx}
Joost Schymkowitz, Jesper Borg, Francois Stricher, Robby Nys, Frederic Rousseau, and Luis Serrano.
\newblock The foldx web server: an online force field.
\newblock {\em Nucleic acids research}, 33(suppl\_2):W382--W388, 2005.

\bibitem{xue2016prodigy}
Li~C Xue, Jo{\~a}o~Pglm Rodrigues, Panagiotis~L Kastritis, Alexandre~Mjj Bonvin, and Anna Vangone.
\newblock Prodigy: a web server for predicting the binding affinity of protein--protein complexes.
\newblock {\em Bioinformatics}, 32(23):3676--3678, 2016.

\bibitem{bank1971protein}
Protein~Data Bank.
\newblock Protein data bank.
\newblock {\em Nature New Biol}, 233(223):10--1038, 1971.

\bibitem{bennett2025atomically}
Nathaniel~R Bennett, Joseph~L Watson, Robert~J Ragotte, Andrew~J Borst, D{\'e}Jena{\'e}~L See, Connor Weidle, Riti Biswas, Yutong Yu, Ellen~L Shrock, Russell Ault, et~al.
\newblock Atomically accurate de novo design of antibodies with rfdiffusion.
\newblock {\em bioRxiv}, pages 2024--03, 2025.

\bibitem{fang2024helixfold}
Xiaomin Fang, Jie Gao, Jing Hu, Lihang Liu, Yang Xue, Xiaonan Zhang, and Kunrui Zhu.
\newblock Helixfold-multimer: Elevating protein complex structure prediction to new heights.
\newblock {\em arXiv preprint arXiv:2404.10260}, 2024.

\bibitem{notin2023proteingym}
Pascal Notin, Aaron~W Kollasch, Daniel Ritter, Lood van Niekerk, Steffanie Paul, Hansen Spinner, Nathan Rollins, Ada Shaw, Ruben Weitzman, Jonathan Frazer, et~al.
\newblock Proteingym: Large-scale benchmarks for protein design and fitness prediction.
\newblock {\em bioRxiv}, 2023.

\bibitem{shanker2024unsupervised}
Varun~R Shanker, Theodora~UJ Bruun, Brian~L Hie, and Peter~S Kim.
\newblock Unsupervised evolution of protein and antibody complexes with a structure-informed language model.
\newblock {\em Science}, 385(6704):46--53, 2024.

\bibitem{zambaldi2024novo}
Vinicius Zambaldi, David La, Alexander~E Chu, Harshnira Patani, Amy~E Danson, Tristan~OC Kwan, Thomas Frerix, Rosalia~G Schneider, David Saxton, Ashok Thillaisundaram, et~al.
\newblock De novo design of high-affinity protein binders with alphaproteo.
\newblock {\em arXiv preprint arXiv:2409.08022}, 2024.

\bibitem{watson2023novo}
Joseph~L Watson, David Juergens, Nathaniel~R Bennett, Brian~L Trippe, Jason Yim, Helen~E Eisenach, Woody Ahern, Andrew~J Borst, Robert~J Ragotte, Lukas~F Milles, et~al.
\newblock De novo design of protein structure and function with rfdiffusion.
\newblock {\em Nature}, 620(7976):1089--1100, 2023.

\bibitem{barlow2018flex}
Kyle~A Barlow, Shane Ó~Conchúir, Samuel Thompson, Pooja Suresh, James~E Lucas, Markus Heinonen, and Tanja Kortemme.
\newblock Flex ddg: Rosetta ensemble-based estimation of changes in protein--protein binding affinity upon mutation.
\newblock {\em The Journal of Physical Chemistry B}, 122(21):5389--5399, 2018.

\bibitem{larson2020x}
Eric~T Larson, Debra~L Brennan, Eugene~R Hickey, Raj Ganesan, Rachel Kroe-Barrett, and Neil~A Farrow.
\newblock X-ray crystal structure localizes the mechanism of inhibition of an il-36r antagonist monoclonal antibody to interaction with ig1 and ig2 extra cellular domains.
\newblock {\em Protein Science}, 29(7):1679--1686, 2020.

\bibitem{beukenhorst2022influenza}
Anna~L Beukenhorst, Jacopo Frallicciardi, Clarissa~M Koch, Jaco~M Klap, Angela Phillips, Michael~M Desai, Kanin Wichapong, Gerry~AF Nicolaes, Wouter Koudstaal, Galit Alter, et~al.
\newblock The influenza hemagglutinin stem antibody cr9114: Evidence for a narrow evolutionary path towards universal protection.
\newblock {\em Frontiers in Virology}, 2:1049134, 2022.

\bibitem{ekiert2009antibody}
Damian~C Ekiert, Gira Bhabha, Marc-Andr{\'e} Elsliger, Robert~HE Friesen, Mandy Jongeneelen, Mark Throsby, Jaap Goudsmit, and Ian~A Wilson.
\newblock Antibody recognition of a highly conserved influenza virus epitope.
\newblock {\em Science}, 324(5924):246--251, 2009.

\bibitem{phillips2021binding}
Angela~M Phillips, Katherine~R Lawrence, Alief Moulana, Thomas Dupic, Jeffrey Chang, Milo~S Johnson, Ivana Cvijovic, Thierry Mora, Aleksandra~M Walczak, and Michael~M Desai.
\newblock Binding affinity landscapes constrain the evolution of broadly neutralizing anti-influenza antibodies.
\newblock {\em Elife}, 10:e71393, 2021.

\bibitem{lin2022language}
Zeming Lin, Halil Akin, Roshan Rao, Brian Hie, Zhongkai Zhu, Wenting Lu, Nikita Smetanin, Allan dos Santos~Costa, Maryam Fazel-Zarandi, Tom Sercu, Sal Candido, et~al.
\newblock Language models of protein sequences at the scale of evolution enable accurate structure prediction.
\newblock {\em bioRxiv}, 2022.

\end{thebibliography}

\end{document}